\documentclass[proof]{WileyASNA-v1}

\usepackage{graphicx}
\usepackage{amsmath}
\usepackage{amssymb}
\usepackage{bm}
\usepackage{tensor}
\usepackage[makeroom]{cancel}
\usepackage{needspace}
\usepackage{verbatim}
\usepackage{float}
\usepackage{nicefrac}
\newcommand{\RB}{\mathbb{R}}

\newcommand{\im}{\mathrm{i}}
\newcommand{\FC}{F}
\newcommand{\HO}{\Omega}
\newcommand{\ho}{\omega}

\newcommand{\hoc}{\tilde{q}}

\newcommand{\Lag}{L}

\newcommand{\Hf}{H}

\newcommand{\HCd}{\mathcal{\Hf}}
\newcommand{\FCd}{\mathcal{\FC}}
\newcommand{\LCd}{\mathcal{\Lag}}

\newcommand{\rmi}{\im}

\newcommand{\dd}{\,\mathrm{d}}

\newcommand{\onehalf}{{\textstyle\frac{1}{2}}}

\newcommand{\onethird}{{\textstyle\frac{1}{3}}}
\newcommand{\quarter}{{\textstyle\frac{1}{4}}}

\newcommand{\oneeights}{{\textstyle\frac{1}{8}}}

\newcommand{\iquarter}{{\textstyle\frac{\rmi}{4}}}
\newcommand{\onetwelfths}{{\textstyle\frac{1}{12}}}
\newcommand{\must}{\stackrel{!}{=}}

\newcommand{\pfrac}[2]{\frac{\partial{#1}}{\partial{#2}}}

\newcommand{\ppfrac}[3]{\frac{\partial^{2}{#1}}{\partial{#2}\partial{#3}}}

\newcommand{\pvarfrac}[2]{\frac{\delta{#1}}{\delta{#2}}}

\newcommand{\detpartial}[2]{\left| \pfrac{#1}{#2} \right|}
\newcommand{\eref}[1]{Eq.~(\ref{#1})}

\newcommand{\sref}[1]{Section~\ref{#1}}

\articletype{Article Type}%
\received{18 October 2023}
\revised {}
\accepted{}

\begin{document}

\title{On CCGG, the De Donder-Weyl Hamiltonian formulation of canonical gauge gravity} 

\author[1]{David Vasak*}
\author[1]{Johannes Kirsch}
\author[1,2]{Armin van de Venn}
\author[1,2]{Vladimir Denk}
\author[1,2]{J\"urgen Struckmeier} 

\authormark{David Vasak \textsc{et al}}

\address[1]{Frankfurt Institute for Advanced Studies (FIAS), Ruth-Moufang-Strasse~1, 60438 Frankfurt am Main, Germany}
\address[2]{Goethe Universit\"at, Max-von-Laue-Strasse 1, 60438 Frankfurt am Main, Germany}

\corres{Presented at the MAGIC meeting, Praia do Rosa, Santa Catarina, Brasil, 05 -- 09 March 2023\\
*\email{vasak@fias.uni-frankfurt.de} }

\abstract{
This short paper gives a brief overview of the manifestly covariant canonical gauge gravity (CCGG) that is rooted in the De Donder-Weyl Hamiltonian formulation of relativistic field theories,
and the proven methodology of the canonical transformation theory.
That framework derives, from a few basic physical and mathematical assumptions, equations describing generic matter and gravity dynamics with the spin connection emerging as a Yang Mills-type gauge field.
While the interaction of any matter field with spacetime is fixed just by the transformation property of that field, a concrete gravity ansatz is introduced by the choice of the free (kinetic) gravity Hamiltonian.
The key elements of this approach are discussed and its implications for particle dynamics and cosmology presented.
New insights:
Anomalous Pauli coupling of spinors to curvature and torsion of spacetime,
spacetime with (A)dS ground state, inertia, torsion and geometrical vacuum energy,
Zero-energy balance of the Universe leading to a vanishing cosmological constant and torsional dark energy.
}


\maketitle
\section{Introduction} \label{sec:Intro}

General Relativity (GR), based on the Einstein-Hilbert Lagrangian that is torsion-free and linear in the Ricci scalar, has explained a wealth of phenomena on the solar scale and beyond,
and hence became the standard gravity ansatz in astrophysics and cosmology.
However, with increasing knowledge from new observations, GR turned out to lack the predictive power for explaining large structure evolution and the dynamics of the Universe.
As a remedy, auxiliary ``Dark gadgets'' have been added to that theory but those have not been understood yet.
Ad hoc and ``trial and error''-based approaches to modifications of the particle and/or gravity models are still unsatisfactory and often inconsistent.

\medskip
Can a consistent and mathematically sound theory of gravity be \emph{derived} from a limited set of evidence-based assumptions?
This is a long standing question, and in this talk a confirmative answer will be given.

\medskip
We introduce the framework of the Covariant Canonical Gauge Gravity (CCGG), illuminate its content, and discuss some novel interesting findings.
The basis of that framework is the rigorous math of the field-theoretical, canonical transformation theory in the abstract, manifestly covariant De Donder-Weyl (DW)  Hamiltonian formalism~\citep{dedonder30,weyl35,strRei12}.
The result is a generic theory of gravity~\citep{struckmeier17a} that unambiguously fixes  the couplings of matter with a dynamical spacetime.
After specifying the free DW Hamiltonians of matter (scalar, vector and spin-$\nicefrac{1}{2}$ fields) and gravity, the canonical field equations are derived.
With the necessary quadratic modification of the linear Einstein-Cartan gravity, spacetime is endowed with inertia and torsion, and the canonical framework yields Poisson-like equations of motion for curvature and torsion with fermionic source terms, and a local energy conservation equation in the form of an extended, Einstein-type field equation.

\medskip
As there is ``no free meal'', the requirement of mathematical rigor comes at the cost of calculation complexity.
It is now, of course, reasonable to ask, what the payoff is of that approach.
To motivate the reader we list upfront a few key advantages and indicate supporting literature: 
 \begin{itemize}
 \item \emph{Math guidelines:} Application of established framework requires non-degenerate Lagrangians of gravity and 
 matter which enforce~\citep{Benisty:2018ufz} a quadratic extension of the Einstein-Cartan theory supporting renormalizability~\citep{stelle77}.
 \item \emph{Symmetry leads:} No ad hoc recipe (``partial to covariant``) is needed to fix the coupling matter-gravity.
 The couplings are derived in the framework~\citep{struckmeier17a} just from the transformation properties of fields, universally for \emph{any form of action}~\citep{struckvasak15}
\item \emph{Physics input:} The free (uncoupled, kinetic) Hamiltonians for gravity and matter are selected to align with phenomenology and the mathematical requirement of regularity.
This leads to quadratic-linear gravity with torsion, and quadratic Dirac Gasiorowicz formulation~\citep{gasiorowicz66} of spin-$\nicefrac{1}{2}$ dynamics.
For spin-0 and spin-1 fields the Klein-Gordon and Proca-Maxwell theories are applied, respectively~\citep{struckmeier21a}.
\item \emph{Metric-affine Hamiltonian formulation:} Derivatives of the independent fields, vierbein representing the metric and connection representing parallel transport, are replaced by momentum fields giving first-order PDEs. This not only avoids the problem of boundary terms in the Lagrangian, but also the trap of Ostrogradsky instabilities \citep{Ostrogradsky:1850fid,Woodard:2015zca,pittphilsci15932}.
\item \emph{Novel view of spacetime:} Spacetime acquires ``inertia'' from the quadratic Riemann-Cartan curvature concomitant, and additional dynamics from torsion~\citep{struckmeier17a}
\item \emph{Anomalous spinor-gravity coupling:} The non-degenerate Dirac Hamiltonian gives after gauging a novel spinor--curvature interaction with an emergent length as coupling constant~\citep{struckmeier21a}.
\end{itemize}

This presentation is intended to give a high-level overview of the CCGG approach in order to spread the news to a wider community.
For the sake of brevity stipulated by conference schedule we cannot go into much detail but explicit calculations can be found in Ref.~\citep{vasak23}.

We stress that some of the results coincide with those discussed in works that have pursued similar ideas in the Lagrangian formalism.
This includes the so called Poincar\`{e} Gauge Theory (PGT) with a linear Einstein-Cartan ansatz as introduced by Utiyama, Sciama and Kibble~\citep{utiyama56, sciama62, kibble67}, and later developed further by Hehl~\citep{hehl76} and others.

\section{Covariant canonical transformation theory} \label{sec:covcantrafotheory}
In order to understand the framework of CCGG we first set the stage by listing the key underlying mathematical and physical assumptions:
\begin{itemize}
  \item Spacetime is the four-dimensional orthonormal Lorentzian base manifold in the frame bundle (PFB) with fibers that correspond to representations of the Lorentz group SO($1,3$).
  \item The Principle of General Relativity is naturally implemented as the diffeomorphism invariance of the base manifold, and Lorentz invariance of the attached frames, i.e.\ as invariance w.r.t. the symmetry group SO($1,3$)$\times$Diff($M$).
  \item The system consists of matter fields embedded in a curved spacetime that are sections on the tangent or spinor bundle, and its dynamics derives from the Hamiltonian principle.
  \item The total Lagrangian density is thus a functional of matter fields  with definite transformation properties with respect to a representation of the symmetry group~SO($1,3$)$\times$Diff($M$).
  \item That Lagrangian must be non-degenerate for ensuring the existence of a unique De Donder-Weyl (DW) Hamiltonian.
\end{itemize}
Based on these postulates, the canonical transformation theory provides a rigorous guidance to enforcing invariance of the action integral w.r.t. any local symmetry (Lie) group acting on the fields.
The form of free Lagrangian or Hamiltonian densities for all matter and spacetime fields is an independent input, though, that must satisfy the above requirements but also be empirically reasonable.

\subsection{The De Donder-Weyl Hamiltonian field theory} \label{sec:DWHFT}
This manifestly covariant ``multisymplectic'' formalism treats all four dimensions on equal footing, unlike the traditional approach where just the time component assumes a unique role for dynamics. 
It assumes formally the existence of a non-degenerate Lagrangian density $\tilde{\LCd} := \sqrt{-g}\,{\LCd}$ enabling the existence of a corresponding DW Hamiltonian via a Legendre transform involving covariant momentum fields. The factor $\sqrt{-g}$ that transforms the Lagrangian scalar into a scalar density is the invariant volume element necessary to convert the action into a world scalar.

For illustration we consider the example of a real scalar field $\phi$.
With comma denoting the partial derivative with respect to the spacetime coordinate $x$, the canonical momentum field is defined as
\begin{equation*}
 \tilde{\pi}^\mu := \pfrac{\tilde{\LCd}(\phi,\phi_{,\mu})}{\phi_{,\mu}}.
\end{equation*}
The Legendre transform of the Lagrangian density gives the De Donder-Weyl (DW) Hamiltonian
\begin{equation}\label{def:Hsimple}
\tilde{\HCd}(\phi,\tilde{\pi}^\mu) := \tilde{\pi}^\mu\,\phi_{,\mu} - \tilde{\LCd}.
\end{equation}
Then the action can be expressed as
\begin{equation}
S =  \int_{V}\tilde{\LCd}\,\dd^4x
=
\int_{V}\left(\tilde{\pi}^\mu\,\phi_{,\mu} - \tilde{\HCd}\right)\,\dd^4x.
\end{equation}
The dynamics of the field results from the Hamiltonian principle, i.e.\ by variation of the action integral w.r.t. the now independent conjugate fields $\phi$ and $\tilde{\pi}^\mu$.
This obviously means:
\begin{subequations}
 \begin{align}\label{eq:canonicalequations0}
\phi_{,\nu}&=\pfrac{\tilde{\HCd}}{\tilde{\pi}^\nu} \\
\tilde{\pi}\indices{^\nu_{,\nu}}&=-\pfrac{\tilde{\HCd}}{\phi}.
\end{align}
\end{subequations}
These \emph{canonical equations} are first-order partial differential equations.
Notice that if $\tilde{\HCd}$ would not depend on the momentum, the conjugate field $\phi$ would be cyclic, resulting in the constraint equation $\phi_{,\mu} = 0.$
Hence if a DW Hamiltonian is independent of a momentum field, the latter assumes the role of a Lagrange multiplier.
The canonical equations are equivalent to the Euler-Lagrange equations when the momentum can be expressed by the field derivatives.
That is possible only if the Hamiltonian is at least quadratic in the momentum field, which, on the other hand, is achievable iff the Lagrangian is non-degenerate.

A remark is appropriate at this point.
The DW Hamiltonian $\HCd$ is, unlike the standard Hamiltonian in point mechanics, \emph{not} the energy of the system.
That energy is rather expressed by the canonical energy-momentum tensor,
\begin{equation*}
{\theta}\indices{_\mu^\nu}
:=\pfrac{{\LCd}_{\mathrm{}}}{\phi_{,\nu}}\,\phi_{,\mu}
-\delta_\mu^\nu\,{\LCd}_{\mathrm{}},
\end{equation*}
derived from Noether's theorem.
However, the trace of that tensor gives
\begin{equation*}
{\theta}\indices{_\mu^\mu} \equiv {\theta} = {\HCd} - 3{\LCd},
\end{equation*}
a covariant version of the relation $\theta = \rho - 3p$ for a homogeneous fluid in its co-moving frame.

\paragraph{Aside: Covariant Hamiltonian dynamics}
The DW Hamiltonian formalism can also be used to define Poisson brackets and the corresponding Hamiltonian dynamics in field theory,
providing a route to covariant quantization.
The modification relative to the standard approach based on time derivatives, is that the covariant Poisson brackets are, due to the vector property of the momentum fields, a co-vector defined as:
\begin{equation*}
\left\lbrace F,G \right\rbrace_\mu := \int\!\! \dd^4 z \,\left(\pvarfrac{F}{\phi(z)} \, \pvarfrac{G}{\tilde{\pi}^\mu(z)} - \pvarfrac{F}{\tilde{\pi}^\mu(z)} \, \pvarfrac{G}{\phi(z)} \right).
\end{equation*}
For the covariant version of Hamilton's equations of motion we need the Hamiltonian world scalar ${H} := \int \! \dd^4 y \, \tilde{\HCd}(y)$.
A straightforward calculation using the canonical equations gives indeed:
\begin{equation*}
\{\phi(x),H\}_\mu \!= \!\int\!\! \dd^4 z \left(
\pvarfrac{\phi(x)}{\phi(z)}  \pvarfrac{H}{\tilde{\pi}^\mu(z)} -
\pvarfrac{\phi(x)}{\tilde{\pi}^\mu(z)}  \pvarfrac{H}{\phi(z)} \right)
= \phi_{,\mu}.
\end{equation*}
This remarkable structure will not be used in the following but is foreseen for future work on covariant quantization.

\subsection{Local canonical transformations in curved spacetimes} \label{sec:CTT}
In the following a brief overview is given on the covariant canonical transformation theory for matter fields embedded in a curved spacetime, described by
a principal frame bundle called Lorentzian manifold.
There we assume to have a basis manifold $M$ representing spacetime as a collection of points, with the generic metric tensor field $g$,
that defines the (in general curved) geometry of $M$.
The points $p \in M$ are assigned a ``coordinate label'' $x$ on every environment $p \in U \subseteq M$ via a collection of  bijective maps $x: U \mapsto x(U) \subset \RB^4$.
The unification of all environments covers $M$ and is called atlas.
The choice of that atlas is not unique but free up to arbitrary diffeomorphisms.
The tangent space of the bundle is equipped with a global Minkowski metric $\eta_{ij}$ and a basis
consisting of the orthonormal system $\{ e^i\}$.
The components of the basis vectors on the base manifold, $e\indices{^i_\nu}$, are called vierbeins (aka tetrads).
Because of the relation
\begin{equation}
g_{\mu\nu} \equiv \eta_{ij}\, e\indices{^i_\mu}\,e\indices{^j_\nu}
\end{equation}
they are also known as the ``square root'' of the metric.
The bases are fixed up to (orthochronous) Lorentz transformations, i.e.\ (a subgroup of) the symmetry group SO($1,3$).
Hence at each point $p$ of the manifold we have locally the freedom to independently choose a map and a vierbein system,
i.e.\ we have the \emph{gauge} freedom with respect to chart transitions $x \mapsto X(x)$ and Lorentz transformations $e\indices{^i_\nu}(p) \mapsto E\indices{^I_\nu}(p)$
via $\Lambda\indices{^I_i}(p)$.
Physics, though, must not depend on that arbitrariness.
This is the Principle of General Relativity that translates here into the requirement that the action be invariant under the symmetry group SO($1,3$)$\times$Diff($M$).

Matter fields are sections on the tangent space of the bundle, and the geometry of spacetime is represented by the vierbein fields.
To illustrate how the transformation properties of these fields under that symmetry group drive the process of canonical transformations, we again, for the sake of simplicity, consider just a real scalar field.
For the system $\varphi(x), e\indices{^i_\nu}(x)$ the transformation from the original frame, denoted by small letters and indices, to the transformed system, denoted by capital letters and indices, is:
\begin{subequations}
\begin{align}
 \varphi(x) &\mapsto \Phi(X) = \varphi(x) \\
 e\indices{^i_\nu}(x) &\mapsto E\indices{^I_\mu}(X) = \Lambda\indices{^I_i}(x)\,e\indices{^i_\nu}(x)\,\pfrac{x^\nu}{X^\mu}.\label{def:trafotetrad}
\end{align}
\end{subequations}
For the dynamics to be invariant under that transformation, the change of the action integral must be restricted to a boundary term on which the fields are fixed.
This is equivalent to the following integrand condition for the Lagrangian density:
\begin{align}
&\tilde{\LCd}^\prime\left(\Phi,\pfrac{\Phi}{X^\nu},E\indices{^I_\mu},\pfrac{E\indices{^I_\mu}}{X^\nu},X\right)\detpartial{X}{x}\\
&\qquad \must\,
\tilde{\LCd}\left(\varphi,\pfrac{\varphi}{x^\nu},e\indices{^i_\mu},\pfrac{e\indices{^i_\mu}}{x^\nu},x\right)-\pfrac{\tilde{\FCd}^\nu}{x^\nu}.\nonumber
\end{align}
In the DW Hamiltonian formulation (with the momentum field $\tilde{k}\indices{_i^{\mu\nu}}$ conjugate to vierbein) we express the Lagrangian density by the equivalent (reverse) Legendre transform:
\begin{align}
&\left[\tilde{\Pi}^\nu\,\pfrac{\Phi}{X^\nu} -
\tilde{K}\indices{_I^{\mu\nu}}\,\pfrac{E\indices{^I_\mu}}{X^\nu} -
\tilde{\HCd}^\prime\left(\Phi,\tilde{\Pi}^\nu,E\indices{^I_\mu},\tilde{K}\indices{_I^{\mu\nu}},X\right)\right]
\detpartial{X}{x}\nonumber \\
&\must\,
\tilde{\pi}^\nu\,\pfrac{\phi}{x^\nu} -
\tilde{k}\indices{_i^{\mu\nu}}\,\pfrac{e\indices{^i_\mu}}{x^\nu} -
\tilde{\HCd}\left(\phi,\tilde{\pi}^\nu,e\indices{^i_\mu},\tilde{k}\indices{_i^{\mu\nu}},x\right)
-\pfrac{\tilde{\FCd}^\nu}{x^\nu}.
\end{align}
While the first two terms on both sides of this equation display the appropriate transformation property, the Hamiltonian density must obviously satisfy
the so called canonical transformation rule
\begin{align}\label{eq:Htransform}
&\tilde{\HCd}^\prime\left(\Phi,\tilde{\Pi}^\nu,E\indices{^I_\mu},\tilde{K}\indices{_I^{\
mu\nu}},X\right)\detpartial{X}{x} \\
& \qquad  =
\tilde{\HCd}\left(\phi,\tilde{\pi}^\nu,e\indices{^i_\mu},\tilde{k}\indices{_i^{\
mu\nu}},x\right)
+\left.\pfrac{\tilde{\FCd}^\nu}{x^\nu}\right|_{\text{expl}}.\nonumber
\end{align}
The vector density $\tilde{\FCd}^\nu$ is the key ingredient of the canonical transformation theory called \emph{generating function}.
Its design must be such that it implements the transformation property of involved matter fields with respect to a given local symmetry.
For relativistic field  theories, it exists only in four versions depending on the four possible combinations of original and transformed conjugate fields.
Here, for enforcing symmetry with respect to the local SO($1,3$)$\times$Diff($M$) field transformations, we choose for convenience the generating function
(called type 3) that depends on the original momenta and the transformed fields.
While the scalar field does not change upon the above symmetry transformation, the vierbein transforms as a vector with respect to both indices.
This is reflected in the specific form of the generating function:
\begin{equation}
\tilde{\FCd}_3^\nu\left(\Phi,\tilde{\pi}^\nu, E\indices{^I_\mu},
\tilde{k}\indices{_i^{\mu\nu}},x\right)
=-\tilde{\pi}^\nu\,\Phi-\tilde{k}\indices{_i^{\beta\nu}}\,\Lambda\indices{^i_I}\,E\indices{^I_\alpha}\,\pfrac{X^\alpha}{x^\beta}.
\end{equation}

\section{Generic gauge gravity} \label{sec:gaugefields}

Obviously, the explicit derivative of that generating function in \eref{eq:Htransform} acts on the spacetime-dependent transformation matrices $\detpartial{X}{x}$ and $\Lambda\indices{^i_I}$:
\begin{align}\label{HLorentzandchart}
\left.\pfrac{\FCd_3^\nu}{x^\nu} \right|_{\text{expl}}&=
-\tilde{k}\indices{_i^{\beta\nu}}\pfrac{}{x^\nu}\left(\Lambda\indices{^i_I}\pfrac{X^\alpha}{x^\beta}\right)E\indices{_\alpha^I}\\
&=-\tilde{k}\indices{_i^{(\beta\nu)}}\pfrac{}{x^\nu}\left(\Lambda\indices{^i_I}\pfrac{X^\alpha}{x^\beta}\right)E\indices{_\alpha^I}
-\tilde{k}\indices{_i^{[\beta\nu]}}\pfrac{\Lambda\indices{^i_I}}{x^\nu}\pfrac{X^\alpha}{x^\beta}E\indices{_\alpha^I}.\nonumber
\end{align}
It does not vanish reflecting the lack of the required local symmetry of the original Lagrangian and the corresponding Hamiltonian densities.
Using the partial derivative of the transformation law,~\eref{def:trafotetrad}, the terms $-\tilde{k}\indices{_i^{(\beta\nu)}}\pfrac{e\indices{_\beta^i}}{x^\nu}$ and $\tilde{K}\indices{_I^{(\beta\nu)}} \pfrac{E\indices{_\beta^I}}{X^\nu}\detpartial{X}{x}$ can be combined with similar terms in \eref{eq:Htransform} to give
\begin{align}\label{F3derivative2concr}
&-\pfrac{\tilde{\pi}^\alpha}{x^{\alpha}}\,\varphi
-\pfrac{\tilde{k}\indices{_i^{[\mu\alpha]}}}{x^{\alpha}}\,e\indices{_\mu^i}
-\tilde{\HCd}\left(\varphi,\tilde{\pi}^\nu,e\indices{_\mu^i},\tilde{k}\indices{_i^{\mu\nu}},x\right)\\
&-\left[\tilde{\Pi}^\nu\pfrac{\Phi}{X^{\nu}}+\tilde{K}\indices{_I^{[\mu\nu]}}\pfrac{E\indices{_\mu^I}}{X^\nu}-
\tilde{\HCd}^\prime \left(\Phi,\tilde{\Pi}^\nu,E\indices{_\mu^I},\tilde{K}\indices{_I^{\mu\nu}},X\right)\right]\detpartial{X}{x}\nonumber\\
&\qquad=\tilde{k}\indices{_i^{[\beta\nu]}}\Lambda\indices{^i_I}\pfrac{\Lambda\indices{^I_j}}{x^\nu} e\indices{_\beta^j}.\nonumber
\end{align}
The remaining term on the right-hand side of \eref{F3derivative2concr} contains the spacetime-dependent Lorentz transformation coefficients $\Lambda\indices{^I_j}(x)$.
The only way to re-establish the invariance of the system dynamics is to introduce a "counter term" whose transformation rule absorbs the symmetry-breaking term proportional
to $\partial\Lambda\indices{^I_j}/\partial x^\nu$. 
That new term called gauge Hamiltonian must thus transform as
\begin{equation}\label{eq:inv-cond}
\tilde{\HCd}_{\mathrm{Gau}}^{\prime}\,\detpartial{X}{x}-\tilde{\HCd}_{\mathrm{Gau}}=
\tilde{k}\indices{_i^{[\mu\nu]}}\Lambda\indices{^i_I}\pfrac{\Lambda\indices{^I_j}}{x^\nu}\,e\indices{_\mu^j}.
\end{equation}
The gauge Hamiltonian is chosen such that the {free} indices $i,j,\nu$ of $\Lambda\indices{^i_I}\partial\Lambda\indices{^I_j}/\partial x^\nu$ are exactly matched:
\begin{equation}\label{g-ham1}
\tilde{\HCd}_{\mathrm{Gau}}=-\tilde{k}\indices{_i^{[\mu\nu]}}\,\ho\indices{^i_{j\nu}}\,e\indices{_\mu^j}.
\end{equation}
Thereby the newly introduced gauge field~$\ho\indices{^i_{j\nu}}$ must retain its form when transformed, hence:
\begin{equation}\label{g-ham2}
\tilde{\HCd}_{\mathrm{Gau}}^{\prime}=-\tilde{K}\indices{_I^{[\mu\nu]}}\,\HO\indices{^I_{J\nu}}\,E\indices{^J_\mu}.
\end{equation}
$\HO\indices{^I_{J\nu}}$ is the transformed gauge field, and from the transformation relation~\eqref{eq:inv-cond} it follows that that transformation must be inhomogeneous:
\begin{equation}\label{omegatransform1}
\ho\indices{^i_{j\nu}}=\Lambda\indices{^i_I}\,\HO\indices{^I_{J\alpha}}\,\Lambda\indices{^J_j}\,\pfrac{X^\alpha}{x^\nu}
+\Lambda\indices{^i_I}\,\pfrac{\Lambda\indices{^I_j}}{x^{\nu}}.
\end{equation}
As this is exactly the transformation property of the spin connection, the gauge field can be identified with the spin connection.
(Notice that the sign of the gauge field was chosen to support this identification.)

The covariant canonical transformation theory thus \emph{derives} gravity as a Yang-Mills type gauge theory wielding
four independent dynamical gravitational fields: the vierbein, $e\indices{^i_\mu}$, representing the geometry, the gauge field spin connection, $\omega\indices{^i_{j\nu}}$, defining parallel transport,
and the respective conjugate momentum fields, $\tilde{k}\indices{_i^{\mu\nu}}$ and $\hoc\indices{_i^{j\mu\nu}}$, defined as:
\begin{equation}
 \tilde{k}\indices{_i^{\mu\nu}} \equiv k\indices{_i^{\mu\nu}}\varepsilon :=
\pfrac{\tilde{\LCd}_{\mathrm{\mathrm{tot}}}}{e\indices{^i_{\mu,\nu}}}   \qquad
\hoc\indices{_i^{j\alpha\beta}} \equiv q\indices{_i^{j\alpha\beta}}\varepsilon
:=
\pfrac{\tilde{\LCd}_{\mathrm{\mathrm{tot}}}}{\omega\indices{^i_{j\alpha,\beta}}}
 \end{equation}
 with $\varepsilon := \det e\indices{^k_\beta} \equiv \sqrt{-\det g_{\mu\nu}}$.

The resulting action integral is a world scalar, and the integrand is form-invariant under the transformation group~SO($1,3$)$\times$Diff($M$):
\begin{align}\label{def:actionintegral0}
S_0 &=
\int_{V}\tilde{\LCd}_{\mathrm{\mathrm{tot}}}\,\dd^4x \\
&=
\int_{V}\left(\tilde{k}\indices{_i^{\mu\nu}}\,S\indices{^i_{\mu\nu}}
+\onehalf \,
\hoc\indices{_i^{j\mu\nu}}\,R\indices{^i_{j\mu\nu}}-\tilde{\HCd}_{\mathrm{Gr}}+
\tilde{\LCd}_{\mathrm{\mathrm{matter}}}\right)\dd^4x.\nonumber
\end{align}
Compared to \eref{def:Hsimple}, the field derivatives (``velocities'') of the vierbein and the connection have in the gauging procedure miraculously morphed into covariant field strengths, namely torsion of spacetime and Riemann-Cartan curvature, respectively, defined as:
 \begin{subequations}
\begin{align}
S\indices{^i_{\mu\nu}} :=& \,\onehalf \left(
\pfrac{e\indices{^i_\mu}}{x^\nu} - \pfrac{e\indices{^i_\nu}}{x^\mu}
+ \omega\indices{^i_{j\nu}} \, e\indices{^j_\mu} - \omega\indices{^i_{j\mu}} \,
e\indices{^i_\nu} \right)\\ 
 \equiv& \,e\indices{^i_\lambda}\,S\indices{^\lambda_{\mu\nu}} =  e\indices{^i_\lambda}\,\gamma\indices{^\lambda_{[\mu\nu]}}
\nonumber \\
R\indices{^i_{j\mu\nu}} :=& \,\pfrac{\omega\indices{^i_{j\nu}}}{x^\mu} -
\pfrac{\omega\indices{^i_{j\mu}}}{x^\nu} +
\omega\indices{^i_{n\mu}}\,\omega\indices{^n_{j\nu}} -
\omega\indices{^i_{n\nu}}\,\omega\indices{^n_{j\mu}} \\
\equiv& \, e\indices{^i_\lambda}\,e\indices{_j^\sigma}\,R\indices{^\lambda_{\sigma\mu\nu}} \nonumber \\
=& \,e\indices{^i_\lambda}\,e\indices{_j^\sigma}\,\left(\pfrac{\gamma\indices{^{\lambda}_{\sigma\nu}}}{x^{\mu}} -
 \, \pfrac{\gamma\indices{^{\lambda}_{\sigma\mu}}}{x^{\nu}} +
 \, \gamma\indices{^{\lambda}_{\delta\mu}}\gamma\indices{^{\delta}_{\sigma\nu}} -
 \, \gamma\indices{^{\lambda}_{\delta\nu}}\gamma\indices{^{\delta}_{\sigma\mu}}\right) \nonumber
\,.
\end{align}
\end{subequations}
This identification is achieved as the expression
\begin{equation} \label{def:gammainomega2}
\gamma\indices{^{\mu}_{\alpha\nu}}
:= e\indices{_{k}^{\mu}}
\left(\pfrac{e\indices{_{\alpha}^{k}}}{x^{\nu}} + \omega\indices{^{k}_{i\nu}} \, e\indices{_{\alpha}^{i}}
\right)
\end{equation}
can be identified with the affine connection.
The proof is straightforward since the transformation law for the affine connection,
\begin{equation} \label{gammatransform}
\Gamma\indices{^{\alpha}_{\nu\beta}} \,=\,\gamma\indices{^{\sigma}_{\eta\mu}}\,
  \pfrac{x^\eta}{X^\nu} \,\pfrac{x^\mu}{X^\beta} \,\pfrac{X^\alpha}{x^\sigma}
  \,-\,\pfrac{x^\eta}{X^\nu}\,\pfrac{x^\mu}{X^\beta}\,\ppfrac{X^\alpha}{x^\mu}{x^\eta},
\end{equation}
derives from the transformation law~\eqref{omegatransform1} of the spin connection.
Notice that here and in the following the affine connection coefficients are not independent fields but just a placeholder for the right-hand side\
of the definition \eref{def:gammainomega2}. 

It is useful for a more compact notation to define a covariant derivative on the frame bundle denoted by ``$;$'', that acts on both the Lorentz and coordinate indices.
Then we can re-write the definition~\eqref{def:gammainomega2} as
\begin{equation}\label{def:vierbeinenpostulat3}
e\indices{_\mu^{i}_{;\nu}}=\pfrac{e\indices{_\mu^i}}{x^\nu}+\omega\indices{^i_k_\nu}\,e\indices{_\mu^k}\,-\,
\gamma\indices{^\alpha_\mu_\nu}\,e\indices{_\alpha^i}\equiv0.
\end{equation}
This is called the Vierbein Postulate, and ensures metric compatibility, i.e.\ the vanishing covariant derivative of the metric and thus the preservation of lenghts and angles,
\begin{equation} \label{def:metricity}
g\indices{_{\mu\nu;\alpha}}(x)=-e\indices{_\mu^i}\,e\indices{_\nu^j}\,\left(\omega\indices{_j_i_\alpha}+\omega\indices{_i_j_\alpha}\right)=0,
\end{equation}
provided the spin connection is anti-symmetric in $ij$ which we shall assume henceforth.

\medskip
The formally introduced gauge field remains an external constraint unless its dynamics is specified via a (``kinetic'') Hamiltonian fixing its vacuum dynamics.
Hence in order to close the system, free gravity Hamiltonian density $\tilde{\HCd}_{\mathrm{Gr}}$ was added in~\eref{def:actionintegral0}.
However, it is important to stress here that the action integral~\eqref{def:actionintegral0} is generic as it has been {derived} exclusively from the transformation properties of the fields without specifying any involved free field Lagrangians or Hamiltonians!
Hence we can conclude that for any gauge gravity, aligned with the very general list of assumptions given above, the following abstract properties apply:
\begin{itemize}
 \item The spin connection coefficients, identified as a Yang-Mills type gauge fields, and metric (vierbeins) are independent fields (metric-affine aka Palatini formalism).
 \item Coupling of matter and gravity is fixed.
 \item Torsion and non-metricity are not a priori excluded.
 \item Momentum fields replace derivatives of fields (``velocities'') giving first-order differential equations.
 \item Quadratic curvature momentum tensor concomitants are \emph{necessary} in the free gravity Hamiltonian due to its postulated non-degeneracy \citep{Benisty:2018ufz}.
 \item Metric compatibility is achieved by setting the gauge field to be anti-symmetric, $\omega\indices{_{(ij)\nu}} = 0$.
 \end{itemize}

\section{Linear-quadratic ansatz for $\tilde{\HCd}_{\mathrm{Gr}}$ } \label{sec:CCGG}

Now beyond the formally derived generic gauge theories of gravity, a specific free gravity ansatz will be considered that satisfies the premises posted above,
but which in addition is also consistent with phenomenology on the solar-scale.
This ensures that the Einstein-Hilbert ansatz, General Relativity, can be recovered as a limit.
In the present formulation the dynamics of free gravity is expressed by a DW Hamiltonian density built from the independent momentum tensor densities canonically conjugate to the independent fields vierbein and connection.
The \emph{CCGG} theory extends General Relativity by quadratic 
concomitants of those momentum fields:
 \begin{align}\label{eq:ham-free-grav}
\tilde{\HCd}_{\mathrm{Gr}}=&\frac{1}{4g_{1}\varepsilon}\tilde{q}\indices{_{i}^{j
\alpha\beta}}
\tilde{q}\indices{_{j}^{i\xi\lambda}}\,
e\indices{^n_\alpha}\,e\indices{^m_\xi}\,\eta_{nm}\,e\indices{^k_\beta}\,e\indices{^l_\lambda}\,\eta_{kl}\\
&+
g_{2}\,\tilde{q}\indices{_{i}^{j\alpha\beta}}\,e\indices{^i_\alpha}\,e\indices{^n_\beta}\,\eta_{nj} \nonumber\\
&+\frac{ {1}}{2g_3\varepsilon}\,\tilde{k}\indices{_{i}^{\alpha\beta}}
\tilde{k}\indices{_{j}^{\xi\lambda}}\,
\,\eta^{ij}e\indices{^n_\alpha}\,e\indices{^m_\xi}\,\eta_{nm}\,e\indices{^k_\beta}\,e\indices{^l_\lambda}\,\eta_{kl}
,\nonumber
\end{align}
The constraint of the affine connection to be symmetric as in General Relativity is dropped here, which introduces torsion of spacetime as an additional degree of freedom of spacetime.
By variation of the action integral~\eqref{def:actionintegral0} we obtain the canonical equations specific for 
$\tilde{\HCd}_{\mathrm{Gr}}$, beyond the generic prototype~\eqref{eq:canonicalequations0}:
\begin{subequations}
 \begin{align}
\pfrac{\tilde{\HCd}_\mathrm{Gr}}{\tilde{k}\indices{_i^{\mu\nu}}}&=S\indices{^i_\mu_\nu}\label{eq:caneq1}\\
\pfrac{\tilde{\HCd}_\mathrm{Gr}}{\hoc\indices{_i^{j\mu\nu}}}&=\onehalf R\indices{^i_{j\nu\mu}}.\label{eq:caneq2}
\end{align}
\end{subequations}
The dependence of $\tilde{\HCd}_{\mathrm{Gr}}$ on the momentum field $\tilde{k}\indices{_{j}^{\xi\lambda}}$ ensures a non-vanishing torsion, and metric compatibility by the implicitly stipulated
anti-symmetry of the spin connection.

Inserting the DW Hamiltonian~\eqref{eq:ham-free-grav} into the canonical equation~\eqref{eq:caneq1} relates the canonical momentum field, conjugate to the vierbein, to the dual ``velocity'' field. 
In the covariant formulation this gives:
\begin{subequations}
 \begin{align}
q\indices{_{i}^{j\alpha\beta}} =& g_1\left(R\indices{_{i}^{j\alpha\beta}} -
\bar{R}\indices{_{i}^{j\alpha\beta}}\right) \\
\bar{R}\indices{^{i}_{j\mu\nu}} :=& -g_2 \left(
e\indices{^i_\mu}\,e\indices{^k_\nu} - e\indices{^i_\nu}\,e\indices{^k_\mu}
\right) \eta_{kj}\,.
\end{align}
\end{subequations}
The non-covariant ``velocity'' $e\indices{^i_{\mu,\nu}}$ is replaced by the Riemann-Cartan curvature tensor multiplied by the deformation parameter $g_1$.
The resulting \emph{canonical} momentum tensor is displaced from the \emph{kinetic} momentum tensor by the maximally symmetric curvature tensor,
\begin{equation}\label{eq:Rmaxsym}
\bar{R}\indices{^j_{i\mu\nu}}=-g_2\left(e\indices{_\mu^j}\,g_{\nu\lambda}-e\indices{_\nu^j}\,g_{\mu\lambda}\right)e\indices{_i^\lambda},
\end{equation}
that is the curvature tensor of the de Sitter (dS with $g_2<0$) or anti-de Sitter (AdS with $g_2>0$) spacetime.
Hence the canonical momentum field describes a deformed geometry relative to the (A)dS ground state of spacetime, and
the constant $g_1$ appears, in analogy to {mass} in classical mechanics, as the {inertia} of spacetime.

\medskip
The second canonical equation~\eqref{eq:caneq2} identifies the momentum field conjugate to the vierbein (i.e.\ metric) with torsion of spacetime:
\begin{equation}
 k\indices{_{j}^{\xi\lambda}} = g_3\,S\indices{_{j}^{\xi\lambda}}.
\end{equation}
Since there is no term linear in $ k\indices{_{j}^{\xi\lambda}}$ in the Hamiltonian, the canonical and kinetic momenta coincide here.

\medskip
The DW Hamiltonian~\eqref{eq:ham-free-grav} thus describes a (metric compatible) spacetime that is endowed with inertia and enriched by torsion.
Notice that with $g_3 = 0$ the spacetime is torsion-free, and $k\indices{_{j}^{\xi\lambda}}$ becomes the corresponding Lagrange multiplier.
With $g_1 = 0$ we had a vanishing momentum, i.e.\ a spacetime without autonomous dynamics.
(That analogy with point mechanics is illustrated in the Aside below.)
And omitting the first and last term gives the Einstein-Hilbert theory.

\paragraph{Aside: Quadratic-linear Hamiltonian in point mechanics}

For illustration on the impact of the linear term in the Hamiltonian we consider a simple example from classical point mechanics, the harmonic oscillator.
Quadratic dependence of the Lagrangians in point mechanics on the velocity $v = \dot{x}$ implies a quadratic dependence of the corresponding Hamiltonian on the canonical momentum $p$.
We thus consider a particle with inertial mass $m$ attached to a spring vibrating around~$x_0$,
extend the Hamiltonian by a linear momentum term, and add a constant energy:
 \begin{equation*}
H(x,p) = \frac{p^2}{2m} +\alpha p + V(x) + E_0, \qquad V(x) = \onehalf k (x-x_0)^2.
\end{equation*}
While the constant does not impact the canonical equations, the linear term does:
\begin{align*}
\dot{x} &= \pfrac{H}{p} = \frac{p}{m} + \alpha \qquad \implies {p = m(\dot{x}-\alpha)} \\
-\dot{p} &= \pfrac{H}{x} = k (x-x_0) \qquad \implies m\ddot{x} = k (x-x_0).
\end{align*}
It shifts the canonical momentum $p$ with respect to the kinetic momentum $m\dot{x}$.
The equation of motion of the harmonic oscillator is not changed, though.
However, after performing the Legendre transformation we observe an emergent constant term in the Lagrangian,
\begin{equation*}
L(x,\dot{x}) = p\dot{x} - H(p(x,\dot{x}),x) = \onehalf m\dot{x}^2 - \alpha m\dot{x} + {\onehalf m\alpha^2} - E_0 - V(x).
\end{equation*}
In order to eliminate the constant from the Lagrangian we set $E_0 \equiv {\onehalf m\alpha^2}$, giving a geometric meaning to the vacuum energy $E_0$ in the Hamiltonian.

\medskip
In the next step we drop the quadratic term in the Lagrangian:
\begin{equation*}
L(x,\dot{x}) = - \alpha m\dot{x} - \onehalf k (x-x_0)^2.
\end{equation*}
The particle has now a constant momentum and is attached to the minimum of the potential:
\begin{equation*}
p = \pfrac{L}{\dot{x}} = \alpha m \qquad 0 \must \dot{p} = -\pfrac{V}{x} = k (x-x_0) \qquad \implies  {x = x_0}.
\end{equation*}
The velocity is not explicitly determined but obviously given by a possible motion of the origin of the potential well.
A linear Lagrangian lets the particle rigidly follow the given potential, and its inertia and ability to absorb kinetic energy is absent, independently of the particle's mass.

\medskip
Recalling the linear Einstein-Hilbert theory suggests a remarkable analogy:
The linear Lagrangian leads to a spacetime that rigidly follows matter.
$\nabla_\mu R^{\nu\mu} \equiv 0$ corresponds to the ``vanishing momentum'' of spacetime, and $\nabla_\mu \theta^{\nu\mu} \must 0$ is the ``potential well'' facilitated by matter.

\section{The CCGG field equation} \label{sec:fieldequation}

Combining the partial derivatives of the canonical equations leads to an extended version of Einstein's field equation:
\begin{align}  \label{def:consistency}
&g_1\,\Big( R^{\alpha\beta\gamma\mu}R\indices{_{\alpha\beta\gamma}^{\nu}}
\!\!-\!\quarter g\indices{^\mu^\nu}R^{\alpha\beta\gamma\xi}  R_{\alpha\beta\gamma\xi} \Big)\\
&\quad+\! {\frac{1}{8\pi G}}\Big(R\indices{^{(\mu\nu)}}\!\!-\!\onehalf
g\indices{^\mu^\nu}\!R
\!-\! {\lambda_0} g\indices{^\mu^\nu} \Big)\nonumber\\
&\quad-2g_3\left(S^{\xi\alpha\mu}S\indices{_\xi_\alpha^\nu}-\onehalf
S^{\mu\alpha\beta}S\indices{^\nu_\alpha_\beta}
-\quarter\,g^{\mu\nu}S_{\xi\alpha\beta}S^{\xi\alpha\beta}\right)=\!\theta^{(\mu
\nu)}.\nonumber
\end{align}
$\theta^{(\mu\nu)}$ is the {symmetric} portion of the canonical stress-energy tensor.
That symmetrization is derived from the canonical equations avoiding the need for any external effort a la Belifante-Rosenfeld~\citep{rosenfeld40}.
Notice that the left-hand side of this equation is derived in complete analogy to the right-hand side, and can thus formally be interpreted as the energy-momentum tensor of space-time.
The gravity ansatz is ``deformed'' by a quadratic term parametrized by $g_1$, the {deformation parameter}.
Remarkably, the appearance of the quadratic invariant in the field equation is in line with Einstein's anticipation that he expressed in his letter to Hermann Weyl, see Fig. 1. 
\begin{figure}[htb]\hspace*{-5mm}\label{fig:letter}
\includegraphics[width=95mm]{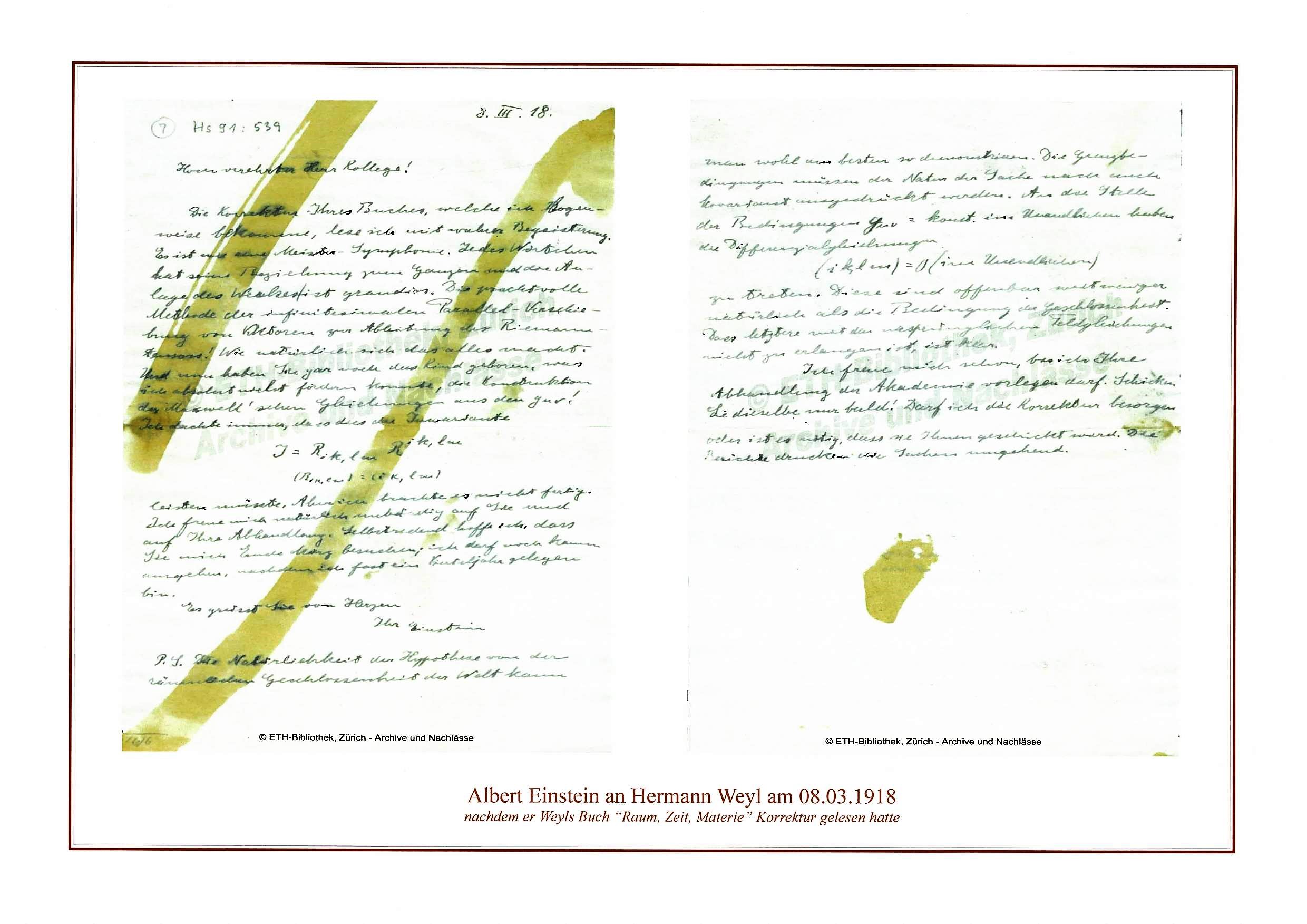}
 \caption{\footnotesize Einstein wrote to Weyl on 8 March 1918, after having reviewed his book ``Space, Time, Matter'': \emph{``... I have always thought that the invariant $I = R_{iklm}\,R^{iklm}$ should make it. But I did not succeed to accomplish that.``} (From the archive of ETH Z\"urich.)}
\end{figure}
The second term corresponds to the Einstein-Cartan formulation, and the third is a torsional contribution to energy-momentum.

\medskip
The weak field limit links the product of the newly introduced coupling constants $g_1$ and $g_2$ to Newton's constant $G$ and to the constant $\lambda_0$,
\begin{subequations}
\begin{align}
3g_2 &\equiv \lambda_0\\
2g_1\,g_2 &\equiv \frac{1}{8\pi G}\equiv
M_{\mathrm{p}}^2,
\end{align}
and the field equation then align with the notation of General Relativity.
These relations imply
\begin{align}
\lambda_0 &= \frac{3M_{\mathrm{p}}^2}{2g_1}.
\end{align}
\end{subequations}
The constant $\lambda_0$ is the \emph{geometric vacuum energy of spacetime} that is {distinct from the vacuum energy of matter}!
Notice, in addition, that the Schwarzschild metric solves the CCGG field equation~\citep{kehm17}, ensuring {consistency with solar scale observations}.

\section{Zero-Energy Universe and the Cosmological Constant}\label{sec:zeroenergyuniverse}

The left-hand side of the field equation~\eqref{def:consistency} is formally the negative canonical
energy-momentum of spacetime, $-\vartheta^{(\mu\nu)} $, the \emph{strain-energy tensor}~\citep{struckmeier18a}.
Hence in CCGG, in analogy to the stress-strain relation in elastic media, the \emph{local zero-energy condition} for the Universe,
\begin{equation}\label{eq:zero}
\vartheta^{\,\mu\nu}+\theta^{\,\mu\nu}=0,
\end{equation}
emerges,  as conjectured by Lorentz and Levi Civita already in 1916/17~\citep{lorentz1916,levi-civita1917}.

\medskip
In order to understand the relation of the geometric cosmological constant $\lambda_0$ with the ``physical'' cosmological constant as introduced by Einstein, we consider three scenarios~\citep{Vasak:2022gps}.
In absence of any matter including vacuum fluctuations the right-hand side of \eref{eq:zero} vanishes.
Spacetime is then expected to settle in its \emph{static} ground state with vanishing momentum fields.
(Graviton fluctuations are neglected here.)
From the canonical equations we thus get
\begin{subequations}
	\begin{align}
		q\indices{_{\,i}^{j\alpha\beta}} &= 0 \implies
		R\indices{^{\,i}_{j\alpha\beta}} = \bar{R}\indices{^{\,i}_{j\alpha\beta}}  \, ; \\
		k\indices{_{\,i}^{\alpha\beta}} &= 0 \implies
		S\indices{_{\,i}^{\alpha\beta}} = 0 \, ,
	\end{align}
\end{subequations}
where $\bar{R}\indices{_{\,i}^{j\alpha\beta}}$ is defined in Eq.~\eqref{eq:Rmaxsym}.
By substituting $\bar{R}\indices{_{\,i}^{j\alpha\beta}}$ for ${R}\indices{_{\,i}^{j\alpha\beta}}$ on the left-hand side of~\eref{def:consistency}, we find
\begin{align*}
	&g_1\,\left(\bar{R}\indices{_\alpha_\beta_\gamma^\nu}\,
 \bar{R}\indices{^\alpha^\beta^\gamma^\mu}
	-\quarter g^{\mu\nu} \bar{R}\indices{_\alpha_\beta_\gamma_\delta}\,\bar{R}\indices{^\alpha^\beta^\gamma^\delta}\right)\\
&\qquad\qquad	+M_{\mathrm{p}}^2 \,\left(\bar{R}^{(\mu\nu)}-\onehalf g^{\mu\nu}\,\bar{R}-\lambda_0 g^{\mu\nu}\right) \nonumber\\
	&\qquad  = \left(\frac{3M_{\mathrm{p}}^2}{2g_1} - \frac{3M_{\mathrm{p}}^2}{2g_1}\right)\, g^{\mu\nu} = 0.\nonumber
\end{align*}
Hence ${\vartheta}^{\mu\nu} =  0$, the strain-energy of spacetime vanishes, and the relation~\eqref{eq:zero} is satisfied.

\medskip
Spacetime with matter in physical vacuum, i.e.\ void of particles but filled with vacuum energy of matter, $\theta_{\mathrm{vac}}$,  is flat.
With a vanishing curvature the zero-energy condition~\eqref{eq:zero} or~\eqref{def:consistency} must still hold~\citep{Vasak:2022gps}.
Hence:
\begin{equation} \label{eq:lambda0}
\lambda_0 = \frac{3M_{\mathrm{p}}^2}{2 g_1} \must
\frac{\theta_{\mathrm{vac}}}{M_{\mathrm{p}}^2}. 
\end{equation}
Resolving for $g_1$ gives the relation
\begin{equation} \label{eq:detg1}
g_1 = \frac{3M_{\mathrm{p}}^4}{2 \theta_{\mathrm{vac}} }
\end{equation}
which determines the value of the deformation parameter $g_1$.
For example from the ''naive`` cutoff calculation in field theory~\citep{weinberg89} $\theta_{\mathrm{vac}} \approx M_{\mathrm{p}}^4$, hence
$g_1 \approx \nicefrac{3}{2}$.

\medskip
For considering the physical spacetime with particles and vacuum energy, we compare the trace of~\eref{def:consistency} with that of Einstein's field equation that includes the physical cosmological constant $\Lambda$.
Taking into account effects of the more complex dynamical geometry and of quantum effects, the trace can be written as
\begin{align} \label{eq:traceCCGG2}
	R  - \frac{g_3}{M_{\mathrm{p}}^2} S^2 + 4\lambda_0  &=
    \bar{R}   +  R_{\mathrm{geom}}  +  R_{\mathrm{quant}} - \frac{g_3}{M_{\mathrm{p}}^2}\,S^2  +  4\lambda_0  \nonumber\\
    &=  -\frac{1}{M_{\mathrm{p}}^2} \left( 4 \theta_{\mathrm{vac}}  +  \theta_{\mathrm{real}}
	 \right).
\end{align}
$\bar{R}$ is the trace-free Levi-Civita Ricci tensor of General Relativity,
$R_{\mathrm{geom}}$ denotes contribution from dynamical spacetime geometry and torsion.
$R_{\mathrm{quant}}$ stands for some yet unspecified graviton vacuum fluctuations. 
This compares directly with the trace of Einstein's field equation with the observed cosmological constant $\Lambda$
and the ``normalized'' stress-energy tensor $\theta_{\mathrm{real}}^{\mu\nu} = \theta^{\mu\nu} - g^{\mu\nu}\,\theta_{\mathrm{vac}}$ that is void of the vacuum energy:
\setlength\belowdisplayskip{0pt}
\begin{equation} \label{eq:traceEH}
	\bar{R} + 4\Lambda  = -\frac{1}{M_{\mathrm{p}}^2} \, \theta_{\mathrm{real}}.
\end{equation}
Combining now Eqs.~\eqref{eq:lambda0}, \eqref{eq:traceCCGG2} and~\eqref{eq:traceEH} implies
\begin{equation}
\quarter\left(R_{\mathrm{geom}}+R_{\mathrm{quant}}-\frac{g_3}{M_{\mathrm{p}}^2}\,S^2\right)\equiv\Lambda\,.
\end{equation}
The cosmological constant $\Lambda$ as observed today is thus a snapshot of a torsional dynamical \emph{dark energy} term~\citep{vasak23}.

\section{Covariant Poisson equations for matter and spacetime}\label{sec:covpoissoneqs}

The remaining canonical equations lead for the involved (mutually non-interacting) matter fields and spacetime dynamics as described by the specific Hamiltonians (Klein-Gordon for the scalar, Maxwell-Proca for the vector, and Gasiorowicz-Dirac for the spinor fields) to the following Poisson-like equations of motion:
  \begin{itemize}
  \item Axial torsion field is totally skew-symmetric and its source is the anti-symmetric portion of
the Dirac stress-energy tensor (overbar = Levi-Civita connection):
        \begin{subequations}
        \begin{align}
         &S\indices{^\alpha^{\nu\mu}} = S^{[\alpha\nu\mu]}\\
 &{ \nabla}_\alpha\, S\indices{^\alpha_{\nu\mu}}=-\frac{1}{M_{\mathrm{p}}^2 +g_3}T\indices{_{\mathrm{D}\,[\nu\mu]}} \\
  &\qquad\qquad\Longleftrightarrow \nonumber\\
&\bar{\nabla}_\alpha \left[ \left(M_{\mathrm{p}}^2 +g_3\right) S\indices{^{\nu\mu}^\alpha}+ {\Sigma}\indices{^{\nu\mu}^\alpha} \right]\\ 
&\qquad\quad=
S\indices{^\nu_{\beta\alpha}}\,{\Sigma}\indices{^\mu^{\beta\alpha}} -
S\indices{^\mu_{\beta\alpha}}\,{\Sigma}\indices{^\nu^{\beta\alpha}}.\nonumber
 \end{align}
\end{subequations}
   \item The curvature field obeys
  \begin{align}
&g_1\left( \bar{\nabla}_\alpha R\indices{^\nu^{\beta\mu\alpha}}
+R\indices{^\xi^{\beta\mu\alpha}}S\indices{^\nu_\xi_\alpha}
-R\indices{^\xi^{\nu\mu\alpha}}S\indices{^\beta_\xi_\alpha}
\right)\\
&\qquad-(M_{\mathrm{p}}^2 +g_3)\,S\indices{^{\nu\beta}^\mu}=\Sigma\indices{^\nu^\beta^\mu}.\nonumber
\end{align}
The source of curvature is the Dirac spin density and the axial torsion!
\item {Real Klein-Gordon field}
 \begin{equation}
g^{\alpha\beta}\bar{\nabla}_\alpha \bar{\nabla}_\beta \varphi+m^2\varphi=0.
\end{equation}
  \item {Real Maxwell-Proca field}
 \begin{equation}
\nabla_\alpha
{F}\indices{^\mu^\alpha}
-2{F}\indices{^\beta^\alpha}S\indices{^\mu_\beta_\alpha}-m^{2}\,a^{\mu}
=
\bar{\nabla}_\alpha
{F}\indices{^\mu^\alpha}
-m^{2}a^{\mu}
\,=0.
\end{equation}

Neither the scalar nor the vector fields ``see'' torsion!

\item In the Gasiorowicz formulation of the Dirac field the Lagrangian is quadratic in the field derivative.
As input for a canonical transformation this leads to anomalous fermion-gravity couplings with an emergent length parameter $\ell$, and novel couplings to curvature and torsion:
\begin{align}
&\Big[\rmi\gamma^\beta \Big(\pfrac{}{x^\beta} 
-\iquarter\ho_{nm\beta}\,\sigma^{nm} \Big)-m \Big] \,\psi \\
&\quad=\onethird {\ell} \bigg[\oneeights\sigma^{\alpha\beta}\sigma^{nm}\,R_{nm\alpha\beta}\nonumber\\
&\qquad+\rmi S\indices{^\nu_{\beta\xi}}\sigma^{\beta\xi}
\left(\pfrac{}{x^\nu}-\iquarter\ho_{nm\nu}\sigma^{nm} \right)\bigg]\psi.\nonumber
\end{align}
In Minkowski geometry ($\ho\indices{^i_{j\beta}}\equiv 0$, $S\indices{^\nu_{\eta\beta}}\equiv 0$) the standard Dirac equation is recovered: {$\ell$ is spurious for free fermions but becomes physical once interactions are introduced!}
If in curved spacetimes the torsion is neglected, on the other hand, then the spin connection reduces to the Levi-Civita connection $\bar{\ho}_{nm\beta}$, so called rotation coefficients.
This gives an anomalous \emph{curvature-dependent mass correction}:
\begin{equation}
\left[\rmi\gamma^\beta\left(\pfrac{}{x^\beta}-\iquarter\bar{\ho}_{nm\beta}\sigma^{nm}\right)-\left(m+{\onetwelfths \ell R}\right)\right]\psi=0.
\end{equation}
\end{itemize}

\section{Cosmology and torsional dark energy}

Here we investigate how a CCGG based cosmology compares with the astronomical observations and the underlying Concordance model.
The model of the Universe is based on the FLRW metric with the scale factor of the expanding 3D-space, ${a(t)}$,
and the spatial constant curvature parameter $K$:
\begin{align*}
	ds^2 = dt^2-a^2(t) \left[\frac{dr^2}{1-K r^2}+
	r^2(d\theta ^2+\sin^2\theta d\phi ^2) \right].
\end{align*}
This metric is then used to calculate the Levi-Civita portion of the curvature.
In standard cosmology based on General Relativity, the so called $\Lambda$CDM model, this is then plugged into the Einstein equation with the cosmological constant $\Lambda$.
The matter content of the Universe is thereby approximated by ideal ``co-moving'' fluids with the densities $\rho_i(t)$ and pressures $p_i(t)$ for baryonic and dark matter ($i=\,$m) and for radiation ($i=\,$r), with the barotropic equations of state (EOS):
\begin{equation} \label{EOSgeneric}
	p_i = \omega_i\, \rho_i.
\end{equation}
 This gives the well known Friedman-Lema\^{\i}tre equations for $a(t)$:
\begin{subequations}
	\begin{align}
		{\left(\frac{\dot{a}}{a}\right)}^2+\frac{K}{a^2}-\frac{1}{3}\Lambda &= \frac{8\pi G}{3} \sum_{i=m,r} \rho_i
		\label{eq:f1a} \\
	    \frac{\ddot{a}}{a} - \frac{1}{3}\Lambda  &= -\frac{4\pi G}{3} \sum_{i=m,r} \left(\rho_i + 3p_i\right).
		\label{eq:f2a}
	\end{align}
\end{subequations}
Here the dot denotes time derivatives with respect to the universal time $t$, e.g. $\dot{a} \equiv da/dt$.
Under the assumptions, the fluids are inert and non-interacting, the combination of equations~\eqref{eq:f1a}, \eqref{eq:f2a} and \eqref{EOSgeneric} fixes the dynamics of each individual fluid:
\begin{equation}
	\dot{\rho}_i = -3\frac{\dot{a}}{a}\left(\rho_i + p_i  \right)
	=-3\frac{\dot{a}}{a}\,\rho_i \left(1 + \omega_i  \right).
\end{equation}
The EOS parameter for pressure-less matter is $\omega_\mathrm{m} = 0$, and for radiation we get $\omega_\mathrm{r} =  \nicefrac{1}{3}$.
The dynamical impact of the spatial curvature and of the cosmological constant emerges thereby too in the form of geometric ``fluids'' with fixed equations of state, namely with $\omega_\mathrm{K} = -\nicefrac{1}{3}$ and
$\omega_\Lambda = -1$.
With the definition of the so called Hubble parameter,
\begin{equation}
	H(a) := \frac{\dot{a}}{a}\,, \label{eq:Hubblefunc}
\end{equation}
the first Friedman equation \eqref{eq:f1a} can then be re-written as
\begin{equation}
 H^2(a) = \frac{8\pi G}{3} \rho = \frac{8\pi G}{3} \sum_{i=m,r,K,\Lambda} \rho_i\,.		\label{eq:FriedmanGR}
\end{equation}
The cosmological parameters  $\Omega_i,\,i=\,$ m,r,$\Lambda$,K, entering the individual contributions to the total energy density $\rho_{\mathrm{}}$, are dimensionless relative-densities introduced according to the conventions of the $\Lambda$CDM model:
\begin{subequations}
	\begin{align}
		\rho_\mathrm{m} &:= \rho_{\mathrm{crit}}\, \Omega_\mathrm{m} \, a^{-3} \label{eq:rm} \\
		\rho_\mathrm{r} &:= \rho_{\mathrm{crit}}\, \Omega_\mathrm{r} \, a^{-4} \label{eq:rr} \\
		\rho_\Lambda &:= \rho _{crit}\, \Omega_{\Lambda} \label{eq:rl} \\
		\rho_\mathrm{K} &:= \rho_{\mathrm{crit}}\, \Omega_\mathrm{K} \, a^{-2} \label{eq:rK}
	\end{align}
with 
	\begin{align}
		\rho_{\mathrm{crit}} &:= \frac{3H_0^2}{8\pi G} \label{eq:rc} \\
		\Omega_{\Lambda} &:= \frac{1}{3} \frac{\Lambda}{H_0^2} \label{eq:Ol} \\
		\Omega_\mathrm{K} &:= -\frac{K}{H_0^2}\,. \label{eq:Ok}
	\end{align}
\end{subequations}
Notice that with $a_0 = a(t_0)$ denoting the present-day (at the time $t = t_0$) scale of the Universe, $H_0 = H(a_0)$ stands for the present-day value of the Hubble parameter.
Utilizing now the dimensionless time $\tau = t H_0$ and setting for the overdot henceforth the derivative $\dd / \dd \tau$,
\eref{eq:FriedmanGR} is recast in terms of the normalized Hubble function as
\begin{equation}
	E^2(a) := \frac{H^2(a)}{H_0^2} 
	= \Omega_\mathrm{m}a^{-3}+\Omega_\mathrm{r} a^{-4}+\Omega_\Lambda+\Omega_\mathrm{K} a^{-2}\,.
	\label{eq:HubbleGR}
\end{equation}
For $a(\tau_0)=a_0=1$ this gives $E(1) = 1$, and thus
\begin{align}
	\Omega_\mathrm{m} +\Omega_\mathrm{r} +\Omega_{\Lambda} +\Omega_\mathrm{K} = 1 \, . \label{eq:sumOm}
\end{align}
With an extended theory of gravity also the complexity of the geometry-related terms in the Friedman equations increase.
For the CCGG ansatz that will be analyzed in the next section.

\subsection{The Friedman Universe revisited}

The Friedman equation~\eqref{eq:HubbleGR} of the $\Lambda$CDM model, seen in the light of the Zero-energy condition~\eqref{eq:zero}, can be expressed as
\begin{equation}
 \rho_{\mathrm{crit}}\left(\hat{\rho}_\mathrm{st} + \hat{\rho}_\mathrm{matter}\right) = 0,
\end{equation}
such that the partial energy densities of matter and spacetime must always stay in balance.
Since we assume for physical reasons that the energy density of matter is always positive, the contribution of the energy density attributed to the dynamics and geometry of spacetime must be negative:
\begin{subequations}
  \begin{align}
    \hat{\rho}_\mathrm{matter}  :=& \,\hat{\rho}_\mathrm{r} + \hat{\rho}_\mathrm{m} \ge 0 \\
    \hat{\rho}_\mathrm{st} :=& \,\hat{\rho}_\mathrm{kin} +\hat{\rho}_\mathrm{K}+\hat{\rho}_{\Lambda} \le 0.
    \end{align}
\end{subequations}
The individual energy density contributions as they appear in~\eref{eq:HubbleGR} can obviously be identified as follows:
\begin{subequations}
\begin{align} 
		\hat{\rho}_\mathrm{kin} &= -E^2(a) \\
		\hat{\rho}_\mathrm{m} &= \,\Omega_\mathrm{m} \, a^{-3} \\
		\hat{\rho}_\mathrm{r} &= \,\Omega_\mathrm{r} \, a^{-4} \\
		\hat{\rho}_\mathrm{\Lambda} &= \,\Omega_\mathrm{\Lambda} \\
		\hat{\rho}_\mathrm{K} &= \,\Omega_\mathrm{K} \, a^{-2}
\end{align}
While the leading negative term is the kinetic energy of spacetime given by the Hubble term, the energy densities associated with spatial curvature and the cosmological constant depend on the value and sign of the parameters K and $\Lambda$.

\medskip
The CCGG equation now extends the $\Lambda$CDM model by terms associated with the quadratic curvature and torsion, $\hat{\rho}_\mathrm{geo}$ and $\hat{\rho}_\mathrm{tor}$, respectively:
\begin{align} 
	     \hat{\rho}_\mathrm{st} :=& \,\hat{\rho}_\mathrm{kin} +\hat{\rho}_\mathrm{K}+\hat{\rho}_{\Lambda}+\hat{\rho}_\mathrm{{geo}}+\hat{\rho}_\mathrm{{tor}}
	\, .
\end{align}
\end{subequations}
(Details of the geometric terms can be found in Refs.~\citep{vandeVenn:2022gvl,Kirsch:2023iwd,vasak23}.
Since $\hat{\rho}_\mathrm{geo}$ vanishes with $g_1 = 0$, and $\hat{\rho}_\mathrm{tor}$ vanishes for $g_3=0$, the $\Lambda$CDM model is recovered with $g_1=g_3=0$.)
According to the discussion of the cosmological constant in~\sref{sec:zeroenergyuniverse} we shall set $\hat{\rho}_{\Lambda} = 0$, and expect a dark energy term to arise from torsion in
the torsional energy density $\hat{\rho}_\mathrm{tor}$. 

\subsection{Check of concept: Torsion can account for dark energy}

That expectation has been checked analytically~\citep{vandeVenn:2022gvl} and numerically~\citep{Kirsch:2023iwd}.
Both calculations are consistent with the following model assumptions:
\begin{itemize}
 \item For the baryonic and dark matter as well as for radiation we take the standard scaling laws.
\item The spacetime contributions are simplified by setting $\Lambda\,$~=~K~=~0.
\item The parameters used are the standard (Planck) parameters plus $g_1$, and in addition $g_3$ and $s_1$ characterizing the torsional dark energy.
Notice that the vacuum energy of matter, $\theta_\mathrm{vac}$, is determined from $g_1$, or vice versa.
\end{itemize}

\medskip
The key numerical results are displayed in~Fig.2. 
 \begin{figure}[!ht]\vspace*{0mm}\hspace*{-4mm}
    \centering
    \begin{tabular}{cc}
      \includegraphics[scale=0.22]{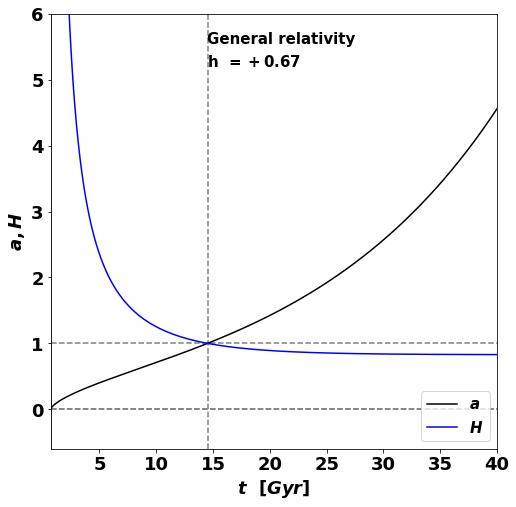} & 
      \includegraphics[scale=0.22]{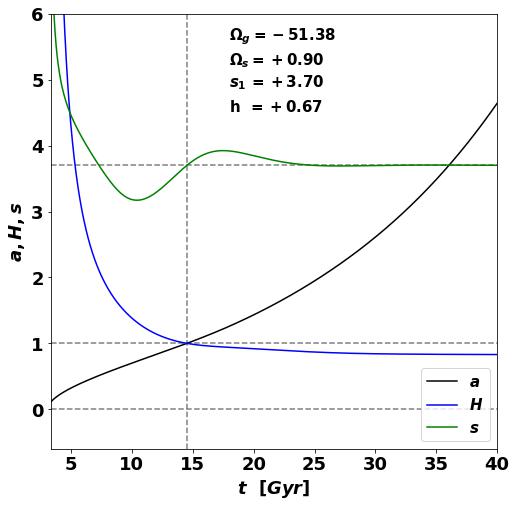} \\
        \includegraphics[scale=0.22]{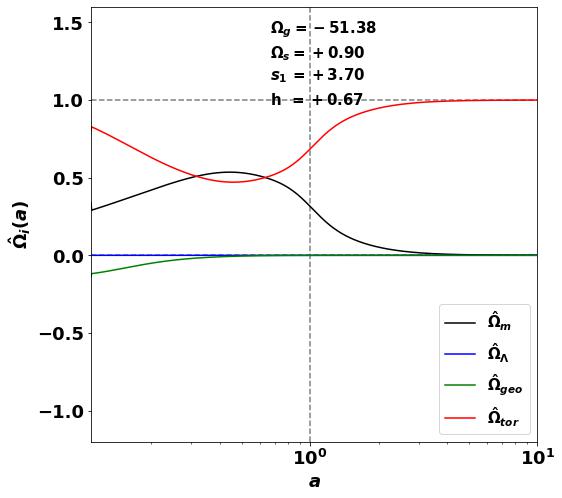}  &
     \includegraphics[scale=0.22]{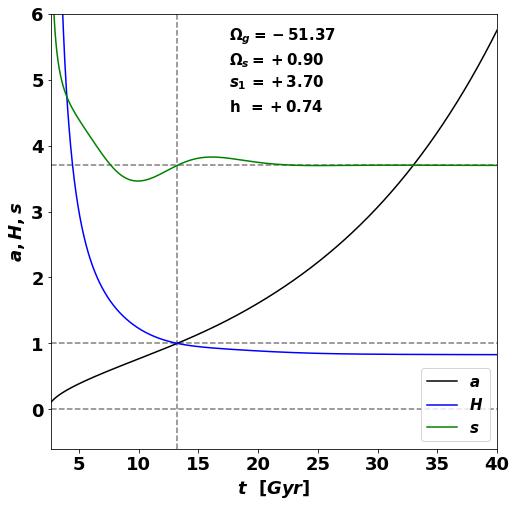} \\
    \end{tabular}
   \caption{\footnotesize
   \emph{Upper left panel:} {The scaling factor $a$ (black line) and the Hubble parameter $H(a)$ (blue line) of General Relativity ($g_1 = g_3 = s_1 = 0$, $\Omega_\Lambda = 0.7$, $h = 0.67$) as a function of the universal time $t$. }
   \emph{Upper right panel:} The scaling factor $a$ (black line), the Hubble parameter $H(a)$ (blue line), and torsion parameter (green line) of the CCGG ansatz with $\Omega_\Lambda = 0$ and $h=0.67$.
   The difference to the evolution of the scale factor in GR is barely visible.
   \emph{Lower left panel:} Evolution of the fractional density parameters $\hat{\Omega}_i(a) = \hat{\rho}_i(a)/E^2(a)$ for the same parameters as in the UR panel. The fractional densities of matter and geometry deformations fade away in an expanding Universe (black and green lines) in contrast to the torsion density (red line) that drives the late acceleration in the dark energy era.
   \emph{Lower right panel:} Same as the UR panel but with $h = 0.74$. }
        \end{figure}\label{fig:2}
We observe that the evolution of the scale parameter in GR with the cosmological constant, and in CCGG cosmology with torsional dark energy, are very similar.
Moreover,  we also find the ratio $\Omega_\mathrm{tor} / \Omega_\mathrm{m}$ to reproduce the ratio $\Omega_\Lambda / \Omega_\mathrm{m} \sim 0.7 / 0.3$.
This resolves both, the cosmological constant and the coincidence problems.
\section{Conclusions}
The key findings of the CCGG ansatz for gravity are summarized as follows:
\begin{itemize}
 \item CCGG provides a novel point of view on gauge gravity beyond the so called Poincar\'e gauge theory. 
 \item Spacetime appears as a dynamical medium with inertia, torsion, and a semi-classical vacuum energy.
 \item Vacuum energies of matter and spacetime cancel: Conjecture of \emph{Zero-Energy-Universe} is confirmed.
 \item A residual torsional  \emph{dark energy}  term is shown to facilitate the  accelerated cosmological expansion.
\item The cosmological constant and coincidence problems are resolved.
 \item A novel length parameter and anomalous couplings of spinors to gravity {emerge}.
\end{itemize}
The present theory is, of course, still far from being finally validated.
The ``torsional dark energy'' hypothesis has to be confirmed.
Albeit the first calculations presented here show promising results, a full-fledged cosmological (MCMC) study vs. all available data is required and under investigation.

The choice of the free gravity Hamiltonian is another field for further studies.
And the implications of the modified structure of the quadratic, Gasiorowicz version of the Dirac equation is yet unexplored as well (length parameter, impact on Lorentz invariance and/or the equivalence principle,
curvature-dependent effective mass, interaction with torsion, inclusion of U(1) and SU(N) symmetries, anomalous magnetic momentum, etc.).

\subsection*{Acknowledgments}
This work has been supported by the Walter Greiner Gesellschaft zur F\"orderung der physikalischen Grundlagenforschung e.V.
DV, JK, AvV and VD thank the Fueck-Stiftung for support.
We thank especially Dr. David Benisty, Dr. Adrian K\"onigstein, Dr. Johannes M\"unch, Dirk Kehm, and Dr. Julia Lienert for their contributions to the development of the CCGG theory.
We are also indebted to colleagues from the local and international community for many valuable discussions:
Prof. Horst St\"ocker (Frankfurt), Prof. Friedrich Hehl (K\"oln), Prof. Gerard ‘t Hofft (Utrecht), Prof. Eduardo Gundelmann (Beer Sheva), Prof. Peter Hess (Mexico City), Dr. Andreas Redelbach (Frankfurt), Dr. Leonid Satarov (Frankfurt), Dr. Frank Antonsen (Copenhagen),  Prof. Pavel Kroupa (Bonn), Prof. Luciano Rezolla (Frankfurt), Prof. Stefan Hofmann (M\"unchen), and Dr. Matthias Hanauske (Frankfurt).



\end{document}